\documentclass{emulateapj}

\usepackage{apjfonts}

\begin{document}

\title{ANN\emph{z}: estimating photometric redshifts using artificial neural networks}  

\author{Adrian A. Collister and Ofer Lahav\altaffilmark{1}}
\affil{Institute of Astronomy, University of Cambridge, Cambridge CB3 0HA, UK}
\email{aac@ast.cam.ac.uk}
\altaffiltext{1}{present address: Department of Physics and Astronomy, University College London, Gower Street, London WC1E 6BT, UK}

\begin{abstract}
We introduce \textsc{ann}\emph{z}, a freely available software package for photometric redshift estimation using Artificial Neural Networks. \textsc{ann}\emph{z} learns the relation between photometry and redshift from an appropriate training set of galaxies for which the redshift is already known. Where a large and representative training set is available \textsc{ann}\emph{z} is a highly competitive tool when compared with traditional template-fitting methods.

The \textsc{ann}\emph{z} package is demonstrated on the Sloan Digital Sky Survey Data Release 1, and for this particular data set the r.m.s. redshift error in the range $0\la{z}\la{0.7}$ is $\sigma_\mathrm{rms} = 0.023$.  Non-ideal conditions (spectroscopic sets which are small, or which are brighter than the photometric set for which redshifts are required) are simulated and the impact on the photometric redshift accuracy assessed.

The package may be freely downloaded from \verb+http://www.ast.cam.ac.uk/~aac+.
\end{abstract}


\section{Introduction}
\label{sec.intro}

In its most general sense, the term \emph{photometric redshift} refers to a redshift estimated using only medium- or broad-band photometry or imaging. Most commonly, photometric redshifts are determined on the basis of galaxies' colours in three or more filters (thus giving a very coarse approximation to the spectral energy distribution, hereafter SED), but they could also be based on other properties which can be derived from images, such as the angular size or concentration index. The method has found successful application to deep-field and wide-field surveys, notably the Hubble Deep Field \citep*[e.g.][]{1999ApJ...513...34F}, and the Sloan Digital Sky Survey \citep{2003AJ....125..580C}.

The most commonly used approach to photometric redshift estimation is the 
\emph{template-matching} technique. This requires a set of `template' SEDs covering a 
range of galaxy types, luminosities and redshifts appropriate to the population for 
which photometric redshifts are required. For a particular target galaxy, the 
photometric redshift is chosen to be the redshift of the most closely matching 
template spectrum; this is usually defined as the template which minimizes the 
$\chi^2$ between the template and actual magnitudes. 

The template spectra are usually derived from a small set of SEDs representing 
different classes of galaxy at redshift $z=0$, which are then manually redshifted to 
give a discrete sampling along the redshift axis (note that this method does not account for \emph{evolution} with redshift). Commonly used template sets are the 
\citet*[CWW;][]{1980ApJS...43..393C} SEDs which are derived observationally, or 
those of \citet{BruzChar93}, derived from population synthesis models. The 
template-matching technique owes its popularity to the very few resources required 
for a basic implementation (i.e. a handful of template SEDs), but the 
accuracy of the technique strongly depends on the extent to which the template 
spectra are representative of the target populations: for example, template SEDs 
derived from observations of low-redshift galaxy populations may be a poor match 
for populations at higher redshifts.

The chances of success can be improved by increasing the number of templates, or by 
more carefully matching the templates to the populations being studied. For example, the spectroscopic catalogue of the Sloan 
Digital Sky Survey (SDSS; \citealt{2000AJ....120.1579Y}) could be used to produce a set of templates which are very well representative of the SDSS photometric catalogue \citep{2003AJ....125..580C}. However, in situations with 
such a large amount of prior redshift information about the sample, the 
template-matching technique is not the best approach: so-called \emph{empirical} 
methods usually offer greater accuracy, as well as being far more efficient.

In essence, empirical photometric redshift methods aim to derive a parametrization 
for the redshift as a function of the photometric parameters. The form of this parametrization is deduced through use of a suitably large 
and representative training set of galaxies for which we have both photometry 
and a precisely known redshift. A simple example is to express the redshift as a 
polynomial in the galaxy colours (e.g. \citealt{1995AJ....110.2655C}; 
\citealt{2000AJ....119.2598S}). The coefficients in the polynomial are varied to 
optimize the fit between the predicted and measured redshift. The photometric 
redshift for the galaxies for which we have no spectroscopy can then be estimated by 
applying the optimized function to the colours of the target galaxy.

Ideally the training set would be a representative subset of the actual photometric target sample (this has the attractive side-effect of nullifying any systematics in the photometry). However, the training set could also be derived from a set of template spectra or from simulated catalogues (e.g. \citealt{vanzella}). The photometry for the training set must be for the same filter set and should have the same noise characteristics as that for the target sample. The trained method can usually only be reliably applied to target galaxies within the ranges of redshift and spectral type adequately sampled by the training set.

In this paper we introduce \textsc{ann}\emph{z}, a software package for 
photometric 
redshift estimation using Artificial Neural Networks (hereafter ANNs) to parametrize 
the redshift-photometry relation. It can be shown (e.g. \citealt{jones90}; 
\citealt{blum91}) that a sufficiently complex ANN is capable of approximating to 
arbitrary accuracy any continuous functional mapping. ANNs have previously found a 
number of applications in astronomy, including morphological classification of 
galaxies (e.g. \citealt{1996MNRAS.283..207L}; \citealt{ball}) star/galaxy 
separation \citep{1996A&AS..117..393B} and object detection \citep[e.g.][]{2000MNRAS.319..700A}. \citet*{firth03} previously demonstrated the feasibility of using ANNs for photometric redshift estimation, and more recently \citet{vanzella} have applied the method to the Hubble Deep Fields.

The layout of this paper is as follows. In \S\ref{sec.anns} Artificial Neural 
Networks are introduced, and the particular methods used by \textsc{ann}\emph{z} are 
explained. 
In \S\ref{sec.sdss} \textsc{ann}\emph{z} is applied to the Sloan Digital Sky Survey. The 
results 
are compared with rival photometric redshift estimators and various extensions to the 
basic technique are explained and illustrated. Finally, less ideal conditions are 
simulated to assess the impact on the accuracy of photometric redshift estimation. 
In \S\ref{sec.conc} the results are summarised, and prospects for the application of 
\textsc{ann}\emph{z} discussed.

\section{Artificial neural networks}
\label{sec.anns}

\begin{figure}
\plotone{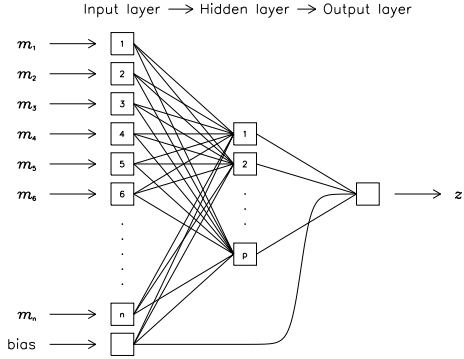}
\caption{\label{fig.photz.arch} A schematic diagram of a multi-layer perceptron, as 
implemented by \textsc{ann}\emph{z}, with input nodes taking, for example, magnitudes $m_i = 
-2.5\log_{10}f_i$ in various filters, a single hidden layer, and a single output node giving, for example, redshift $z$.  The architecture is $n$:$p$:1 in the notation used in this paper.  Each connecting line carries a weight $w_{ij}$. The bias node allows for an additive constant in the network function defined at each node.  More complex networks can have additional hidden layers and/or outputs.}
\end{figure}

\textsc{ann}\emph{z} uses a particular species of ANN known formally as a 
\emph{multi-layer perceptron} (MLP). A MLP consists of a number of layers of 
\emph{nodes} (Fig. \ref{fig.photz.arch}; see e.g. \citealt{bishop}, and references 
therein, for background). The first layer contains the inputs, which in our application to photometric redshift estimation are the magnitudes, $m_i$, of a galaxy in a number of filters (for ease of notation we arrange these in a vector $\mathbf{m} \equiv (m_1,m_2,...,m_n)$). The final layer contains the outputs; we will usually use just one output, the photometric redshift $z_\mathrm{phot}$, but see \S\ref{sec.spec.type} for an example with multiple outputs. Intervening layers are described as \emph{hidden} and there is complete freedom over the number and size of hidden layers used. The nodes in a given layer are connected to all the nodes in adjacent layers. A particular network architecture may be denoted by $N_{\mathrm{in}}$:$N_1$:$N_2$: $\ldots$ :$N_{\mathrm{out}}$ where $N_{\mathrm{in}}$ is the number of input nodes, $N_1$ is the number of nodes in the first hidden layer, and so on.  For example 9:6:1 takes 9 inputs, has 6 nodes in a single hidden layer and gives a single output.

Each connection carries a weight, $w_{ij}$; these comprise the vector of coefficients, $\mathbf{w}$, which are to be optimized. An \emph{activation function}, $g_j(u_j)$, is defined at each node, taking as its argument
\begin{equation}
u_j = \sum_i w_{ij}g_i(u_i), \label{eqn.act}
\end{equation}
where the sum is over all nodes $i$ sending connections to node $j$. The activation functions are typically taken (in analogy to biological neurons) to be sigmoid functions such as $g_j(u_j) = 1/[1 + \exp{(-u_j)}]$, and we follow this approach here. An extra input node -- the bias node -- is automatically included to allow for additive constants in these functions. 

For a particular input vector, the output vector of the network is determined by progressing sequentially through the network layers, from inputs to outputs, calculating the activation of each node (hence this type of neural network is often referred to as a \emph{feed-forward} network).

\subsection{Network training}
\label{sec.nettrain}

Given a suitable training set of galaxies for which we have both photometry, 
$\mathbf{m}$, and a spectroscopic redshift, $z_\mathrm{spec}$, the ANN is trained by minimizing the \emph{cost function}
\begin{equation}
E=\sum_k(z_{\mathrm{phot}}(\mathbf{w},\mathbf{m}_{k}) - z_{\mathrm{spec},k})^2, \label{eqn.cost}
\end{equation}
with respect to the weights, $\mathbf{w}$, where $z_{\mathrm{phot}}(\mathbf{w},\mathbf{m}_{k})$ is the network output for the given input and weight vectors, and the sum is over the galaxies in the training set. To ensure that the weights are \emph{regularized} (i.e. that they do not become too large), an extra quadratic cost term
\begin{equation}
E_\mathrm{w}=\beta\sum_{i,j}w_{ij}^2, \label{eqn.decay}
\end{equation}
is added to equation \ref{eqn.cost}. 

\textsc{ann}\emph{z} uses an iterative quasi-Newton method to perform this minimization. Details of the minimization algorithm and regularization may be found in \citet{bishop} and \citet[][ Appendices]{1996MNRAS.283..207L}.

After each training iteration, the cost function is also evaluated on a separate \emph{validation} set. After a chosen number of training iterations, training terminates and the final weights chosen for the ANN are those from the iteration at which the cost function is minimal on the \emph{validation} set. This is useful to avoid over-fitting to the training set if the training set is small. The trained network may then be presented with previously unseen input vectors, and the outputs computed.

\subsection{Photometric noise}
\label{sec.noise}

In real situations the inputs to the network (e.g. the magnitudes in this case of photometric redshift estimation) will usually have a measurement noise associated with them. We can assess the variance these errors effect in the output using the usual chain-rule approach:

\begin{equation}
\sigma_z^2 = \sum_i{\Big(\frac{\partial{z}}{\partial{m_i}}\Big)^2\sigma_{m_i}^2}, \label{eqn.noise}
\end{equation}
where the sum is over the network inputs.

Given a trained network, the output is an analytic function of the network weights 
and the input vector: $z = z(\mathbf{w},\mathbf{m})$. Provided the activation 
functions, $g_i(u_i)$, are differentiable, the derivatives 
$\partial{z}/\partial{m_i}$ can be obtained through a simple and efficient algorithm 
\citep[][ pp.148--150]{bishop}. This method is used by \textsc{ann}\emph{z} to estimate the 
variance in its photometric redshifts due to the photometric noise.

\subsection{Network variance}
\label{sec.netvar}

Prior to training, \textsc{ann}\emph{z} randomizes the initial values of the weights. Depending on the particular initialization state used, the training process will usually converge to different local minima of the cost function. A simple possibility is to train a number of networks and select one based on the best performance on the validation set. However, this is wasteful of training effort and, in fact, the sub-optimal networks can be used to improve overall accuracy: the mean of the individual outputs of a group of networks (known as a \emph{committee}) will usually be a more accurate estimate for the true redshift than the outputs of any one committee member in isolation.

Using a committee also allows the uncertainty in the output due to the variance in the network weights to be estimated. For a particular target galaxy the photometric redshift prediction should ideally be robust to different intializations of the weight vector. However, it may be the case that the available photometry or training set does not constrain the redshift very well (even for high signal-to-noise photometry, so the error estimated by the method of \S\ref{sec.noise} could be relatively small). These cases are more likely to show a large variance in the output for different initializations of the weight vector, hence using a committee may assist in their identification. \textsc{ann}\emph{z} allows arbitrarily large committees to be used, and estimates the contribution of the network variance to the error in the photometric redshift for each target galaxy.

\subsection{Using the ANN\emph{z} package}

We have made \textsc{ann}\emph{z} available on the WWW at the following address: \verb+http://www.ast.cam.ac.uk/~aac+. Full instructions are provided with the package, but we provide an outline of the procedure here. \textsc{ann}\emph{z} comprises two main programs, \texttt{annz\_train} and \texttt{annz\_test}.

\paragraph{1} When applying \textsc{ann}\emph{z} to any data set for the first time it is strongly recommended that a portion of the available training data be set aside as an evaluation set. This is used as a mock target sample to assess and tune \textsc{ann}\emph{z}'s performance on the data. The evaluation set should therefore be chosen to match the real target sample as closely as possible in terms of its magnitude and colour distributions.

\paragraph{2} The remaining training data should be separated into training and validation sets which are supplied to the \texttt{annz\_train} program along with a description of the required network architecture. This program performs the network training as described in \S \ref{sec.nettrain}. The trained network weights are saved to file.

\paragraph{3} Step {\emph2} may be repeated several times using different network initialisations to obtain a committee of trained networks.

\paragraph{4} The \texttt{annz\_test} program can now be used to apply the trained networks to the target data.

Before applying \textsc{ann}\emph{z} to the actual photometric target sample, the whole procedure should be run several times using the evaluation set as the target data, and varying the parameters of the training (e.g. weight decay, training and validation set sizes, number of networks in the committee) so as to optimize the performance.

\section{Application to SDSS data}

\label{sec.sdss}

The Sloan Digital Sky Survey\footnote{Funding for the creation and distribution of the SDSS Archive has been provided by the Alfred P. Sloan Foundation, the Participating Institutions, the National Aeronautics and Space Administration, the National Science Foundation, the U.S. Department of Energy, the Japanese Monbukagakusho, and the Max Planck Society. The SDSS Web site is \texttt{http://www.sdss.org/} \\ The SDSS is managed by the Astrophysical Research Consortium (ARC) for the Participating Institutions. The Participating Institutions are The University of Chicago, Fermilab, the Institute for Advanced Study, the Japan Participation Group, The Johns Hopkins University, Los Alamos National Laboratory, the Max-Planck-Institute for Astronomy (MPIA), the Max-Planck-Institute for Astrophysics (MPA), New Mexico State University, University of Pittsburgh, Princeton University, the United States Naval Observatory, and the University of Washington.
} 
(SDSS; \citealt{2000AJ....120.1579Y}) combines a large, five-band ($ugriz$) imaging survey with a smaller spectroscopic follow-up survey. This is an ideal situation for the application of \textsc{ann}\emph{z} since the spectroscopic survey represents an excellent training set for the imaging survey.

The selection algorithm for the SDSS spectroscopic survey results in two subsets of the data: a main galaxy catalogue and a luminous red galaxy catalogue \citep[LRG;][]{2001AJ....122.2267E}. The main galaxy catalogue is a flux-limited sample ($r<17.77$) with a median redshift $z=0.104$ \citep{2002AJ....124.1810S}, while the LRG catalogue is flux- and colour-selected to be a very uniform and approximately volume-limited sample (it is volume limited to $z\approx0.4$, but probes out to $z\approx0.6$ at lower completion).

\subsection{Comparison of ANN\emph{z} with other techniques}
\label{sec.compare}

\begin{figure}
\plotone{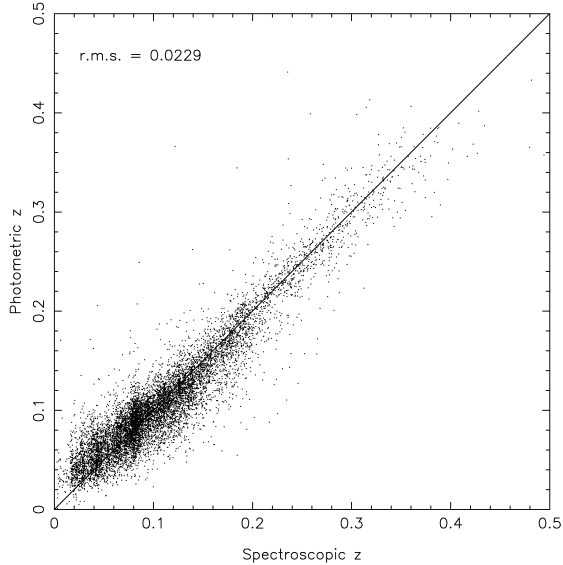}
\caption{\label{fig.edr.5} Spectroscopic \emph{vs.} photometric redshifts for \textsc{ann}\emph{z} applied to 10,000 galaxies randomly selected from the SDSS EDR.}
\end{figure}

\begin{figure}
\plotone{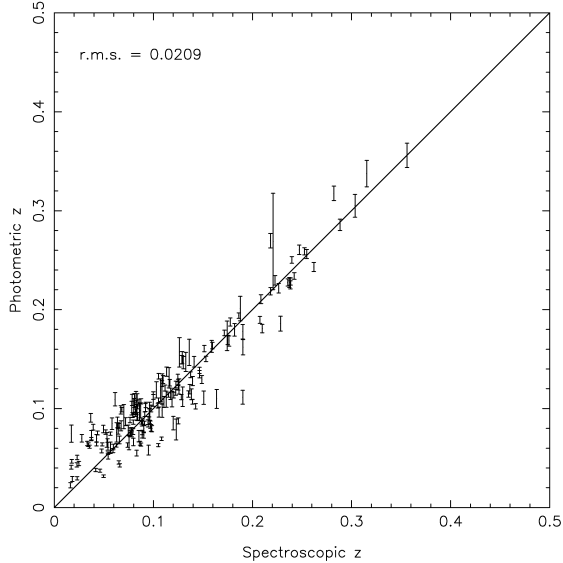}
\caption{\label{fig.edr.errorbars} A subset of 200 galaxies randomly selected from 
the results of Fig. \ref{fig.edr.5}, and with the error bars calculated by 
\textsc{ann}\emph{z} 
shown. These are a combination of contributions from photometric noise (\S\ref{sec.noise}) and network variance (\S\ref{sec.netvar}).}
\end{figure}

The SDSS consortium have themselves applied a range of photometric redshift 
techniques to their commissioning data \citep{2003AJ....125..580C}. Table 
\ref{sdss_results} lists the estimation errors they obtained. This commissioning data 
was made public in the Early Data Release \citep[EDR;][]{2002AJ....123..485S}. In 
order to allow a direct comparison of the accuracy of \textsc{ann}\emph{z} with the methods 
used by \citet{2003AJ....125..580C} we selected the main galaxy and LRG samples from the EDR. From these $\sim$30,000 galaxies we randomly selected training, validation and evaluation sets with respective sizes 15,000, 5,000 and 10,000. The network inputs were the dereddened model magnitudes in each of the five filters and the overall architecture was 5:10:10:1. A committee of five such networks was trained on the training and validation sets, then applied to the evaluation set. Figure \ref{fig.edr.5} shows the \textsc{ann}\emph{z} photometric redshift against the spectroscopic value for each galaxy in the evaluation set. The rms deviation between these is $\sigma_\mathrm{rms} = \sqrt{\langle{}(z_\mathrm{phot} - z_\mathrm{spec})^{2}\rangle{}} = 0.0229$, which compares well with the results in Table \ref{sdss_results}. For clarity the estimated errors on the photometric redshifts are not shown in Fig. \ref{fig.edr.5}. The results for a randomly-selected subset of 200 galaxies are shown with errorbars in Figure \ref{fig.edr.errorbars}. Due to the high quality of the training data in this case, network variance makes only a small contribution and the errors are therefore dominated by the photometric noise.

\begin{deluxetable}{lc}

\tablecaption{Photometric redshift accuracies for the SDSS EDR \label{sdss_results}}

\tablehead{\colhead{Estimation Method} & \colhead{$\sigma_\mathrm{rms}$}}

\startdata
CWW & 0.0666 \\
Bruzual-Charlot & 0.0552 \\
Interpolated & 0.0451 \\
Polynomial & 0.0318 \\
Kd-tree & 0.0254 \\
\textsc{ann}\emph{z} & 0.0229
\enddata

\tablecomments{The first five entries are the photometric redshift accuracies obtained by \citet{2003AJ....125..580C} for the SDSS Early Data Release. The result obtained using \textsc{ann}\emph{z} is appended for comparison.}
\end{deluxetable}

\begin{figure}
\plotone{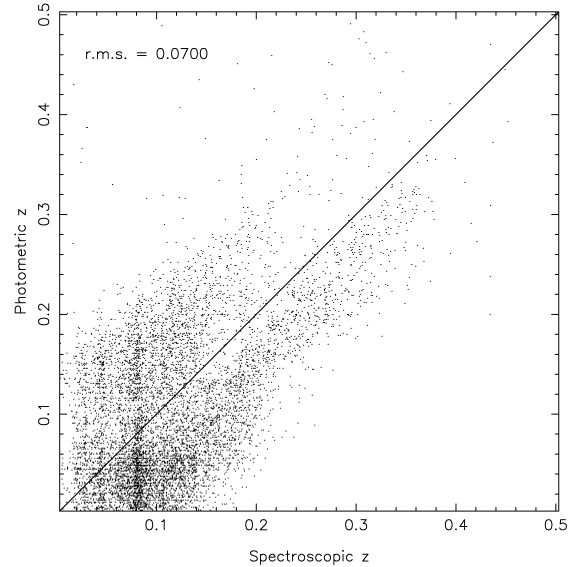}
\caption{Photometric redshift estimation using \textsc{hyperz} with the CWW template SEDs. This uses the same 10,000 galaxy sample as figure \ref{fig.edr.5}. There are obvious systematic deviations, with bands apparent above and below the $z_\mathrm{phot} = z_\mathrm{spec}$ line.}
\label{fig.edr.hyperz}
\end{figure}

\textsc{Hyperz} \citep*{2000A&A...363..476B} is a widely used template-based photometric redshift package. In order to more directly compare \textsc{ann}\emph{z} with the template-matching method, \textsc{hyperz} was applied to the same evaluation set using the CWW template SEDs \citep{1980ApJS...43..393C}. It is clear from the results in Fig. \ref{fig.edr.hyperz} that not only is the rms dispersion in the photometric redshift considerably greater than that for \textsc{ann}\emph{z}, but there are also systematic deviations in the \textsc{hyperz} results. The SDSS consortium obtained similar accuracies to \textsc{hyperz} in their implementation of the basic template-fitting technique (the results labelled \emph{CWW} and \emph{Bruzual-Charlot} in Table \ref{sdss_results} are for the respective template sets). With more sophisticated template-based methods they were able to improve on these errors: the result labelled \emph{Interpolated} was obtained by first tuning the templates using the spectroscopic sample as a training set, then producing a continuous range of templates by interpolating between the tweaked SEDs. However, even ``hybrid'' methods such as this still do not match the accuracy achieved by the purely empirical methods (in the table these are: \emph{Polynomial}, which uses a second-order polynomial as the fitting function, and \emph{Kd-tree}, in which the training set is partitioned in colour-space and a separate second-order polynomial is fitted in each cell).

\subsection{Extensions to the basic method}

In this section more advanced use of \textsc{ann}\emph{z} is demonstrated. These examples use the LRG and main galaxy data from the SDSS Data Release 1 (DR1; \citealt{2003AJ....126.2081A}), split into training, validation and evaluation sets of respective sizes 50,000, 10,000 and 64,175. For these data the photometric redshift accuracy on the evaluation set when using the same basic method as in \S\ref{sec.compare} was $\sigma_\mathrm{rms}= 0.0238$.

\subsubsection{Using additional inputs}
\label{sec.add}

One of the great advantages of empirical photometric redshift methods is the ease with which we can introduce additional observables into our parametrization of the photometric redshift. This is particularly true for \textsc{ann}\emph{z}; we simply add an extra input to our network architecture for each new parameter we wish to consider. \textsc{ann}\emph{z} treats these new inputs in exactly the same way as it does the galaxy magnitudes.

If the additional inputs contain useful information then the ANN will use this to improve the accuracy of its predictions. However, increasing the number of inputs to the ANN generally leads to a reduction in the \emph{generalization} capabilities of the network (that is, its ability to make predictions for data on which it has not been trained). Thus, the inputs should be chosen carefully as non-informative inputs may actually lead to a worsened ANN performance: due to the increased dimensionality of the input space, larger training sets may be required and there will be an increased likelihood of converging to a local rather than the global minimum.

By way of example, the $r$-band 50 and 90 per cent Petrosian flux radii were added as two extra inputs to our ANN. These are the angular radii (concentric with the galaxy brightness distribution) containing the stated fraction of the Petrosian flux, and therefore contain information on the angular size of the galaxy (clearly a strongly distance-dependent property) and the \emph{concentration index} (essentially the steepness of the galaxy brightness profile, which may help break degeneracies in the redshift-colour relationship). Running this extended data set through \textsc{ann}\emph{z} (using a committee of five 7:11:11:1 networks) produced a redshift estimation accuracy of $\sigma_\mathrm{rms} = 0.0230$, an improvement of $\sim$3 per cent compared to the results based only on the magnitudes. In this example the improvement is small (mainly because the training sample already provided excellent redshift information), but it demonstrates well how straightforwardly the extra information could be included for consideration by \textsc{ann}\emph{z}.

\subsubsection{Predicting spectral type}
\label{sec.spec.type}

It is equally straightforward to train \textsc{ann}\emph{z} to make predictions for 
properties other than the redshift. Template-matching photometric redshift techniques 
have the useful side effect of assigning an estimated spectral type to each galaxy, 
in addition to estimating the redshift. \citet{firth03} demonstrated the use of ANNs 
to determine spectral types from broad-band photometry. 

\begin{figure}
\plotone{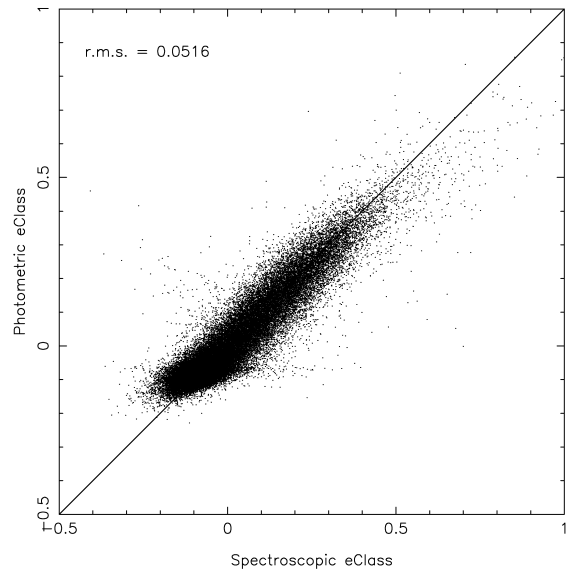}
\caption{\label{fig.eclass} Results from using \textsc{ann}\emph{z} to predict the spectral 
type 
(in the form of the \texttt{eClass} parameter) simultaneously with the redshift for 64,175 galaxies from the SDSS Data Release 1.} 
\end{figure}

The spectroscopic catalogue 
of the SDSS includes a continuous parameter (\texttt{eClass}) indicating spectral 
type which ranges from approximately $-0.5$ (early types) to $1$ (late types). A 
5:10:10:2 network architecture was used to attempt the simultaneous estimation of redshift and \texttt{eClass} from the photometry. The accuracy of the redshift estimation was very slightly poorer, $\sigma_\mathrm{rms} = 0.0241$. The \texttt{eClass} was determined with an rms error of $\sigma_\mathrm{rms} = 0.0516$ (Fig. \ref{fig.eclass}).

\subsection{More realistic conditions}
\label{sec.extrap}

Our example applications to the SDSS above are somewhat idealistic, since we are training and testing on samples with identical redshift, magnitude and galaxy species distributions. Furthermore, our training samples have thus far been very large. In this section less optimal training sets are used to investigate their impact on the photometric redshift accuracy.

\subsubsection{Smaller training sets}

The size of training sample needed will be strongly dependent on the range of redshifts and galaxy types in the target sample. The same evaluation set of 64,175 galaxies was submitted to networks trained on randomly selected samples of (i) 2000 galaxies and (ii) 200 galaxies. In both cases these samples were split equally into the training and validation sets. Committees of five 5:10:10:1 networks were used. 

The photometric redshift accuracies were respectively (i) $\sigma_\mathrm{rms} = 0.0263$ and (ii) $\sigma_\mathrm{rms} = 0.0343$. In the first case the loss of accuracy is small, while the second case demonstrates well the problems associated with small training sets. The rarer classes of object in the target sample (e.g. here, those at high redshift) feature very sparsely (if at all) in the training set and so the network is unable to sensibly deal with these objects when they appear in the testing data. This leads to an increased number of outliers and, potentially, the introduction of systematic errors.

\subsubsection{Biased training sets}
\label{sec.bias}

For increasingly faint targets, acquiring good spectroscopy becomes increasingly difficult and eventually prohibitively expensive; this problem is the primary motivation for photometric redshifts. In practice then, the available spectroscopic training sample is likely to be somewhat brighter on average than the photometric target sets. However, the major stumbling block for empirical photometric redshift estimation techniques is the difficulty in applying them outside of the regions of parameter space which are well sampled by the training data: while the estimator ought to be able to interpolate \emph{within} the training regime, extrapolating beyond is much more problematic. Ideally we would like to be able to train our estimator on bright galaxies, and then confidently apply it to faint galaxies.

We can improve the ANN's prospects by careful pre-selection of the data set. The Luminous Red Galaxies are a very uniform sample with respect to spectral types, since these early-type galaxies show little spectral evolution with redshift; this might be expected to make extrapolation a more manageable task. To assess the effectiveness of \textsc{ann}\emph{z} in this situation the LRG sample was split roughly in half by imposing a magnitude cut at $r=18.5$. The brighter subsample was further divided at random into training and validation sets of size 5000 and 2000 galaxies respectively. A committee of five 5:10:10:1 networks was trained on this data and then applied to the remaining $\sim6000$ LRGs (for which the limiting magnitude is $r\approx19.6$). 

\begin{figure}
\plotone{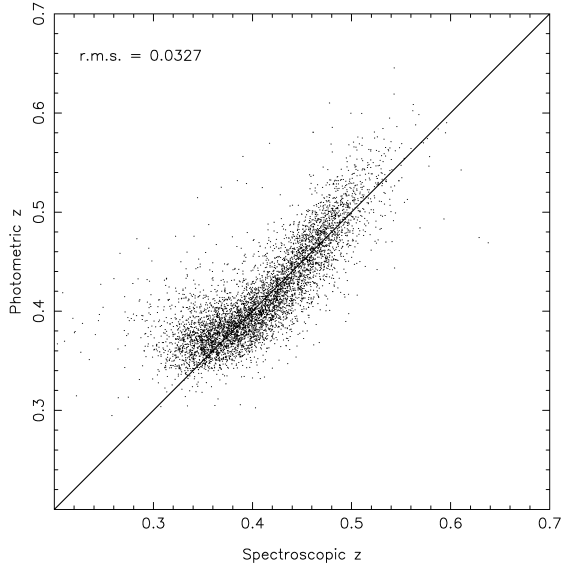}
\caption{\label{fig.lrgs} Results from training networks on LRGs with $r<18.5$, but applied to LRGs with strictly $r>18.5$ (note the change of intercept of the axes). The limiting magnitude for the LRGs is $r\approx19.6$.} 
\end{figure}

The results are shown in Fig. \ref{fig.lrgs}. The overall dispersion is $\sigma_\mathrm{rms} = 0.0327$ which represents only a slight loss of accuracy when compared with results using a LRG training set selected over all magnitudes ($\sigma_\mathrm{rms} = 0.0294$). Thus, in this particular case, \textsc{ann}\emph{z} is able to extrapolate with some success to around a magnitude fainter than is sampled by the training data.

\section{Conclusions}

\label{sec.conc}

In appropriate circumstances, \textsc{ann}\emph{z} is a highly competitive tool for photometric redshift estimation. However, it does rely on the existence of a sufficiently large training set which is representative of the particular populations being studied. The package's utility therefore lies particularly with large photometric surveys such as the SDSS, GOODS \citep{2001AAS...198.2501D} or the VIRMOS-VLT Deep Survey \citep{2003SPIE.4834..173L}, some of which include spectroscopic surveys for subsets of the photometric catalogues (for example, of the eventual 100 million photometric objects which the SDSS expect to catalogue, 1 million will also have spectroscopy, and hence accurate redshifts). 

A major problem for empirical photometric redshift estimators is the difficulty in extrapolating to regions of the input parameter space which are not well sampled by the training data. Care should be taken to match the training data to the target sample as closely as possible in terms of the magnitude and colour distributions of each. Use of an evaluation set is essential when applying \textsc{ann}\emph{z} to a new data set: the good performance demonstrated here on the SDSS data cannot be guaranteed on different data sets.

A potential solution to the problem of obtaining training sets when spectroscopy is difficult to obtain is to use simulated catalogues as training 
data (e.g. \citealt{vanzella}). Since this requires the use of theoretical SEDs it introduces the disadvantages 
of the template-based methods, such as the need for precise calibration. However, the 
ANN approach has advantages over standard template-matching: simulated catalogues can 
contain galaxies representing a large range of complex star formation histories, dust 
extinction models and metallicities etc., giving fully Bayesian statistics, and ANNs 
allow much more flexible weighting to be applied to the filters than is possible with 
the simple $\chi^2$-weighting of standard template-matching.

\acknowledgments

We acknowledge help and advice from Stefano Andreon, Andrew Firth, Rachel Somerville and Elizabeth Stanway. The ANN training program is based on code kindly provided by B.~D.~Ripley. AAC is supported by an Isle of Man Department of Education Postgraduate Studies Grant. OL acknowledges a PPARC Research Senior Fellowship.

Comments on the \textsc{ann}\emph{z} package are welcomed at \verb+aac@ast.cam.ac.uk+.

\end{document}